# Development of wide-field low-energy X-ray imaging detectors for HiZ-GUNDAM

Kazuki Yoshida[a], Daisuke Yonetoku[a], Tatsuya Sawano[a], Hirokazu Ikeda[b], Atsushi Harayama[b], Makoto Arimoto[c], Yasuaki Kagawa[a], Masao Ina[a], Satoshi Hatori[d], Kyo Kume[d], Satoshi Mizushima[d], Takashi Hasegawa[d]

[a]Kanazawa University, Kakuma, Kanazawa, Ishikawa 920-1192, Japan
[b]ISAS/JAXA, 3-1-1, Yoshinodai, Chuo-ku, Sagamihara, Kanagawa 252-5210, Japan
[c]Waseda University, 3-4-1, Ookubo, Shinjuku-ku, Tokyo 169-8555, Japan
[d]The Wakasa Wan Energy Research Center, 64-52-1, Nagatani, Tsuruga, Fukui 914-0192, Japan

## ABSTRACT

We are planning a future gamma-ray burst (GRB) mission HiZ-GUNDAM to probe the early universe beyond the redshift of z > 7. Now we are developing a small prototype model of wide-field low-energy X-ray imaging detectors to observe high-z GRBs, which cover the energy range of 1 – 20 keV. In this paper, we report overview of its prototype system and performance, especially focusing on the characteristics and radiation tolerance of high gain analog ASIC specifically designed to read out small charge signals.

## 1. INTRODUCTION

### 1.1 Scientific Background

It is the most important astronomical issue to reveal the physical condition of the early universe around the epoch just after the end of dark ages and the dawn of formation of astronomical objects. Especially the history and origin of reionization of inter galactic medium, the formation of first stars and first black holes in the universe, and the chemical evolution of heavy elements, should be solved by every conceivable wavelength and method in 2020's astronomy.

The Gamma-Ray Bursts (GRBs) are the most energetic explosions in the early universe, and are one of the most promising tools to probe the early universe. The most distant GRBs, at present, are z = 8.26 of GRB 090423 by spectroscopic observation[1] and z = 9.4 of GRB 090429B by photometric observation[2], respectively. More remote GRBs beyond z > 10 will be detected in near future. Then we can use them as bright light sources to probe the early universe if we can perform rapid follow-up observations while their afterglows are still bright. The first stars (population-III stars, Pop-III stars) would be massive objects, and we expect to detect Pop-III GRBs. Then, we will be enable to investigate the primordial gases surrounding Pop-III stars using their afterglows.

We are planning to a future satellite mission to pioneer the frontier of GRB cosmology, named HiZ-GUNDAM (High-z gamma-ray bursts for unraveling the dark ages mission)[3], which was proposed to ISAS/JAXA's competitive M-class mission in January 2016. HiZ-GUNDAM has two kinds of mission payloads, i.e., a wide-field low-energy X-ray imaging detector to find prompt emission of GRBs, and a near infrared telescope with 30 cm in diameter to find the high-z GRB candidate. We summarize basic information about two mission payloads in Table 1 (slightly updated from Yonetoku et al. 2014).

The wide-field low-energy X-ray imaging detector will play an important role in the observation of electro-magnetic counterpart of gravitational wave (GW) sources, such as GW150914 detected by LIGO on September 14, 2015[4]. It is essential to study some detailed astrophysical properties of GW sources and their environments with electro-magnetic wave. Many telescopes performed follow-up observations for GW 150924, but they failed to detect obvious counterpart for the GW event. This is because the localization accuracy of GW150914 is about 630 deg$^2$ in 90% confidence level[5], and it is too large to survey for narrow field ground telescopes. A couples of LIGO, Virgo[6] and KAGRA[7] will improve the localization accuracy of GW sources to a few 10 deg$^2$[8][9] if they will simultaneously detect the similar GW event. However, for most of weaker GW events, the localization area will be still large, and it is difficult for the ground telescopes to perform follow-up observations effectively. On the other hand, the wide field X-ray observations have some advantages in covering 1~2 steradian and also monitoring the just same region at the moment of GW detection. Thus the localization of X-ray transients as the electro-magnetic counterparts of GW sources will lead the follow-up observations in multi-wavelength. Now we are planning a micro satellite which is named Kanazawa-SAT$^3$[10] and install a wide-field low-energy X-ray imaging detector smaller than one of HiZ-GUNDAM. Kanazawa-SAT$^3$ is a demonstration experiment of X-ray imaging detector aboard HiZ-GUNDAM. In this paper, we introduce a development and performance of this X-ray imaging detector, especially focusing on application specific integrated circuits (ASICs) to readout charge signals.

Table 1. Basic characteristics of mission payloads aboard HiZ-GUNDAM

| Item | Wide-Field X-ray Imaging Detector | Near Infrared Telescope |
|---|---|---|
| Optics | Coded Aperture Mask | Offset Gregorian (focal length 183.5 cm, F=6.1) |
| Energy/Wavelength Range | 1 – 20 keV (goal) <br> 2 – 20 keV (requirement) | 0.5 – 2.5 μm |
| Field of View | > 1 str (Full Coded) | 34 x 34 arcmin$^2$ |
| Localization Accuracy | < 10 arcmin | < 2 arcsec |
| Focal Detector | Silicon Strip Detector (300 μm pitch, 16 mm length) | HyViSi x 1 <br> HgCdTe x 3 |
| Detector Area/Aperture Size | 1,000 cm$^2$ | 30 cm in diameter |
| Sensitivity | $10^{-8}$ erg/cm$^2$/s (1 sec rate trigger) <br> $10^{-9}$ erg/cm$^2$/s (100 sec image trigger) | 0.5 – 0.9 μm : 21.4 mag(AB) <br> 0.9 – 1.5 μm : 21.3 mag(AB) <br> 1.5 – 2.0 μm : 20.9 mag(AB) <br> 2.0 – 2.5 μm : 20.7 mag(AB) <br> (10 min exposure, S/N=10) <br> Simultaneous exposure in 4 bands |
| Alert Message | As soon as possible | Within 30 min <br> Including photometric redshift information |

## 2. WIDE-FIELD LOW-ENERGY X-RAY IMAGING DETECTOR

**2.1 Silicon Strip Detector and Random Coded Aperture Mask**

We selected the 1-dimensional X-ray imaging system with the random coded aperture mask and silicon strip detector (SSD) to achieve the mission requirements summarized in Table 1. We developed several types of SSD which has 64 electrode strips, placed in 300 μm pitches, and are 500 μm thickness (the ratio of photo electric absorption is 98 % for 10 keV X-ray photons)[3][10], and investigated the difference of performance, for example, leakage current, capacitance and ratio of charge separation caused by electrode designs. The total performance is determined by the design of SSD and the noise level of readout ASIC.

The position determination accuracy of the coded-mask imaging system is defined by the geometrical structures, $\theta \sim \tan^{-1}(d/D)$, where d is the strip pitch of the mask and sensor, and D is the distance between mask and detector plane, respectively. We developed a coded aperture mask made of tungsten with 300 μm pitch, and confirmed that position determination accuracy is 5.7 arcminutes with the imaging experiment by using the 5.5-m X-ray beamline in our laboratory at Kanazawa University. The details of imaging experiments are summarized in Sawano et al. 2016.

**2.2 Readout ASIC (ALEX-02)**

We are developing an analog/digital readout system with application specific integrated circuit (ASIC) for the SSDs supported by "Open IP project" promoted by ISAS/JAXA[11]. This ASIC, named "ALEX" series (ASICs for Low Energy X-ray) is specifically designed to readout small charge signals from X-ray with 1 – 20 keV. The first version of the ASIC, "ALEX-01", is reported by Yonetoku et al 2014. The ALEX-01 has a quite large electric gain and achieves the linearity within 2 % deviations between the energy ranges of 1 – 23 keV. But its energy threshold level is 2.3 keV and not satisfied the mission requirement in Table 1 if we define the lower level discrimination threshold to be 4σ of the noise level.

In general, the noise level of semiconductor devices is caused by the electric capacitance, the leakage current, 1/f fluctuation noise and so on. The capacitance of SSD dominates the noise level in our system. According to the SPICE simulation for ALEX circuits, we found that the equivalent noise charge (ENC) is suppressed by having a longer time constant of the shaping amplifier as shown in Figure 1 (left). Thus we developed the second version of ASIC, "ALEX-02", having current source to adjust the time constant (feedback resistor circuit) of fast/slow shaping amplifiers". We can individually adjust their time constants from 0.5 to 5 μs which are longer than the one of ALEX-01. The specifications of the ALEX-02 are summarized in Table 2.

Table 2. Specifications of the ALEX-02

| Chip name | ALEX-02 (ASICs for low energy X-ray ver. 02) |
|---|---|
| Chip size | 8.930 mm × 7.250 mm |
| No. of channel | 64 |
| Pad pitch analog input | 91.2 μm |
| Fabrication and process | X-FAB XH-035 (0.35 μm CMOS) |
| Total gain | 750 mV / fC |
| Peaking time of fast/slow shaping amplifier | 5 μsec (fast), 6 μsec (slow) |
| Power rail | ± 1.65 V (analog and digital), + 3.3 V (digital) |
| Power consumption | ~ 120 mW |

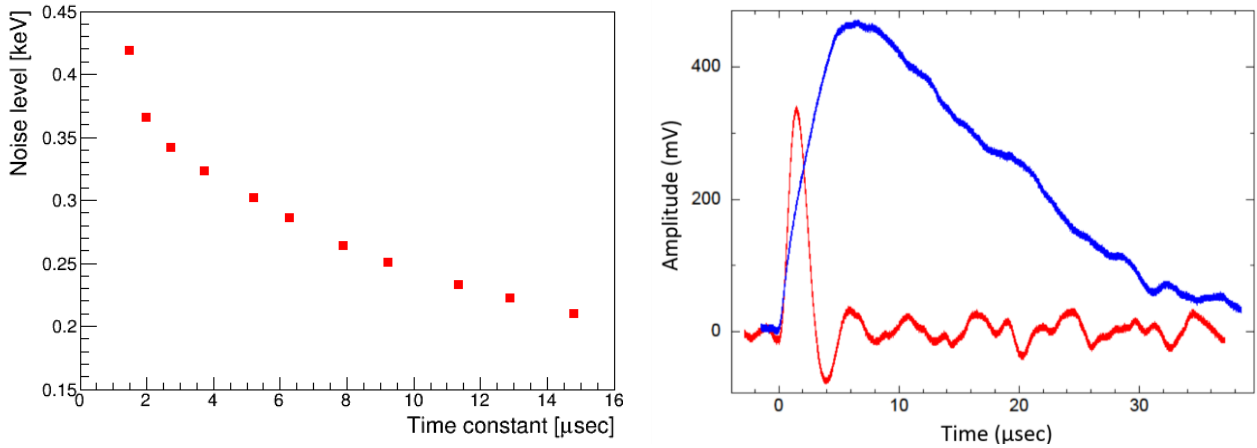

Figure 1. (Left) The noise level of the fast shaping amplifier as a function of shaping time constant calculated by the SPICE simulator. (Right) The observed pulse waveforms of the fast shaping amplifier measured with the ALEX-02 for input charge signal of 10 keV for Si. The red and blue lines are analog output signals for different time constant of 1 μsec (red) and 5 μsec (blue), respectively.

Figure 1 (right) shows two actual waveforms of the fast shaping amplifier with time constants of 1 μsec (red) and 5 μsec (blue). We confirmed that the ENC is suppressed from 138 electrons to 103 electrons which is equivalent to the energy threshold level from 2.0 keV to 1.5 keV, by changing the time constant from 1 μsec to 5 μsec with no-load condition.

At the time of update of ASIC's design, we newly integrated a function of digital-to-analog converters (DAC) to ALEX-02, which adjust voltages for feedback resistors in shaping amplifier circuit, and also the baseline voltages of each 64-channel fast/slow shaping amplifier independently. These DACs are in each channel and there are two types, the 4-bit DAC whose step is ± 40 mV for coarse adjustment and the 5-bit DAC whose step is ± 2 mV for fine adjustment. In addition, the ALEX-02 has five 8-bit DACs, named control DAC, to set the lower/upper threshold, test pulse, VGG and VREF. VGG fixes a feedback resistor of preamplifier circuit in conformity to the leakage current and the capacitance of SSD. VREF configures the offset of ADC in Wilkinson-type ADC. Specifications of control DAC is summarized in Table 3.

Table 3. Specifications of the control DAC

| 8-bit DAC | Voltage/step [mV] | Full range [mV] |
| --- | --- | --- |
| Lower threshold | 2 | ± 256 |
| Upper threshold | 10 | ± 1280 |
| Test pulse | 0.1 | ± 13 |
| VREF | 1 | ± 128 |
| VGG | 5 | ± 640 |

## 3. PERFORMANCE EVALUATION WITH THE ALEX-02 AND SSD

### 3.1 Multi-channel Performance

As shown in Figure 2 (left), we show a bread-board model of SSD and ASIC. The ALEX-02 and SSD[1] are directly connected with wire bonding to reduce any additional stray capacitances. In order to investigate characteristic of the ALEX-02, in the atmospheric temperature of – 20 degrees, we performed several X-ray measurement tests using two radio isotopes and a fluorescent Molybdenum Kα/Kβ lines generated in an X-ray beam line. Figure 2 (right) shows examples of X-ray spectra for a representative arbitral channel with $^{55}$Fe (5.9 keV: green), $^{57}$Co (14.4 keV: blue) and Mo Kα/Kβ lines (17.5 and 19.6 keV: yellow), respectively. The left most structure means a pedestal (baseline noise) level.

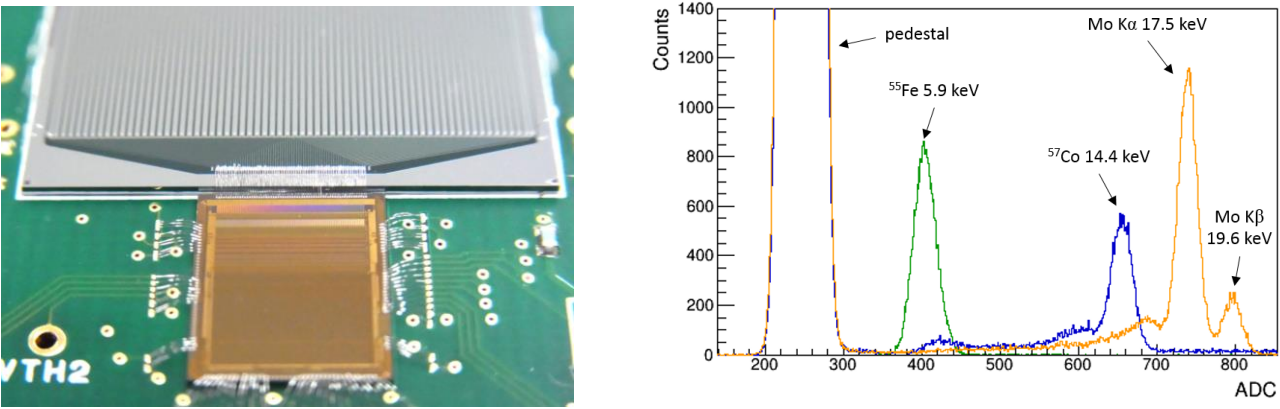

Figure 2. (Left) A photo of ALEX-02 and SSD on a front-end board. (Right) The spectra of $^{55}$Fe, $^{57}$Co and Mo obtained individually with one of the channel.

In the orbital operation, we have to order the energy scale of all channels to determine the single low-energy threshold level. Then, we investigated how we can control the DAC adjustment for baseline voltages and also the gain (or linearity) of signal outputs. First, we individually adjusted the baseline voltages for 64 readout signals with control DACs. Figure 2 (left) shows the distributions of the baseline voltages of fast shaping amplifier outputs. The red and the blue histograms show the baseline distribution before and after the DAC adjustment, respectively. Its standard deviation is drastically changed from 780 eV (red) to 34 eV (blue). We set a single level of trigger threshold voltage for 64 channels in the ALEX-02, then we have systematic error of 34 eV for energy calibration around the threshold level.

Figure 3 (right) shows the distributions of the baseline voltages of slow shaping amplifier outputs. The meaning of colors is similar to Figure 3 (left). Their baselines are also well controlled, and the standard deviation became enough narrow from 2,300 eV to 100 eV. The slow shaping amplifier output varies widely as compared with fast shaping amplifier outputs because of difference of time constant. These pedestal distributions in both shaping amplifiers are well controlled compared with the low-energy threshold of 1 keV as the mission requirement, and much less than the energy resolution as shown in following.

After adjusting the baseline voltages, we measured the linearity of each channel to investigate the uniformity of entire energy range. Figure 4 (left) shows the 64 linearity curves calibrated by 4 energies (i.e. 5.9 keV, 14.4 keV, 17.5 keV and pedestal as zero keV). The red line has the average performance of 64 channels (means of 64 gain slopes and y-intersepts). If we adopt the red line as a representative energy response, the typical 1σ error is equivalent to 5 % (50 eV at 1 keV and 400 eV at 20 keV). These systematic errors are also smaller than the energy resolution and we can accept the average function as energy response.

Next, we estimated the energy resolution for each channel. Here we adopt the average linearity as shown in the previous paragraph (we also analyzed the data with individual linearity, but the results are not strongly affected). Figure

---

[1] SSD has 64 channel strips with 16 mm length. The strip pitch is 300 μm with 200 μm width of aluminum electrodes.

4 (right) shows the distribution of the energy resolution (FWHM) obtained with Mo Kα line (17.5 keV). The mean energy resolution is 1.1 keV and the standard deviation is 0.2 keV. The asymmetry of the histogram may come from the fluctuations of leakage current of SSD and unknown electric disturbance. This should be understood in the future experiments.

Finally, we investigated a realistic energy threshold level of our detector. The energy threshold of electric readout depends on how background count rate we accept. In the energy range of 1 - 20 keV, the cosmic X-ray background (CXB) is the most dominant background irradiation in low earth orbit. Then, setting the several energy threshold level, we estimated the count rates of the electrical noise of readout ASIC (ALEX-02), and compared them with the CXB count rates calculated with the smoothly connected broken-power low model in reported by the Swift observation[12]. After adjusting baseline voltages of all 64 channel readouts, we uniformly set a threshold voltage for all of them and measured net count rate of electric noise (sum of 64 channels). Because of the existence of the several noisy channels, the energy threshold level may be set as 3.5 keV, if we assume the count rate of electric noise are comparable with one of CXB. On the other hand, if we enable to set the performances of all 64 channels to be the best characteristic, we may achieve the energy threshold level of 1.5 keV which is about 30 % levels of the CXB count rate. Therefore, we recognize that it is a crucial issue how to realize the best performance for all readout channel stably.

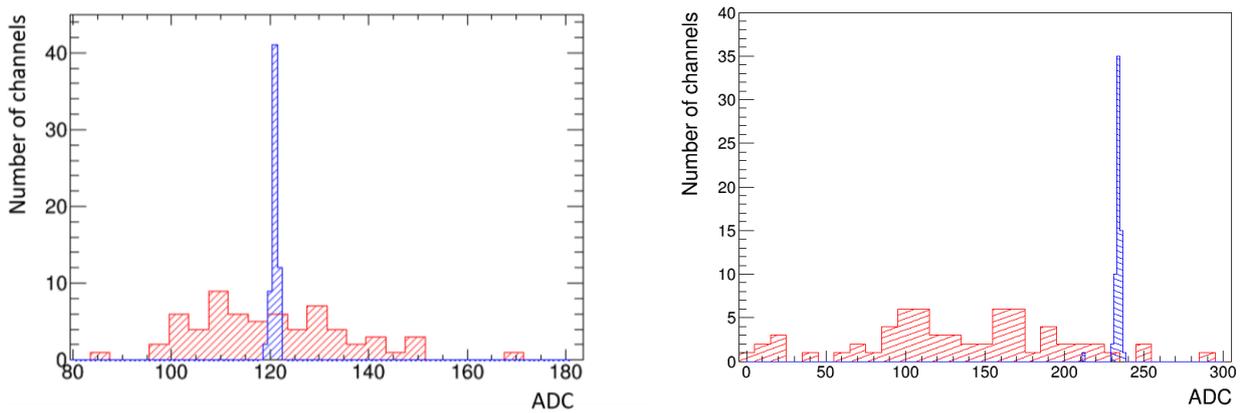

Figure 3. The red and blue histograms show the distribution of baseline voltage before and after adjustment with DAC. The left and right panels show the distribution of fast and slow shaping amplifier output, respectively.

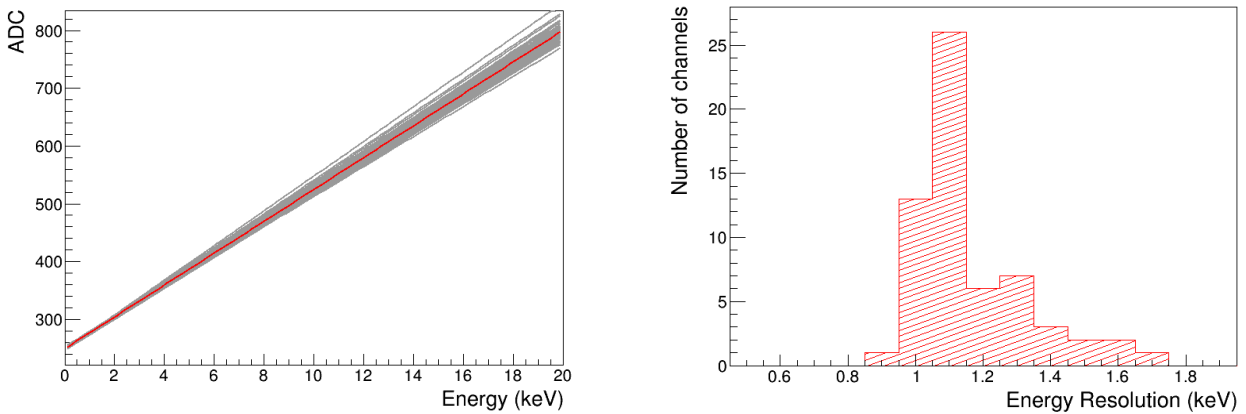

Figure 4. (Left) The linearity of output pulse height (ADC) as a function of X-ray energy after applying baseline voltage adjustment. (Right) The distribution of the energy resolutions (FWHM) of 64 individual channels measured with 17.5 keV of fluorescent Mo Kα line. The mean energy resolution is estimated as 1.1 keV (FWHM).

**3.2 The best Performance**

In this experiment, we used the test SSD which has various width of the electrodes[10]. Figure 5 shows the $^{55}$Fe spectrum obtained with one of the channel that electrode width[2] is 150 μm. The green line is a raw data and the red line is a data subtracted the common mode noise from the raw data. The blue lines are the best fit at Mn Kα and Mn Kβ and the FWHM energy resolution is 0.8 keV. Furthermore, we confirmed the threshold level is 1.9 keV if we define the lower level discrimination threshold to be 4σ of the noise level. This threshold level is lower than the one of the ALEX-01 with no load condition. We will optimize the parameters of the preamplifier circuit such as a trans-conductance of the charge sensitive amplifier so as to reduce the noise level to the requirement.

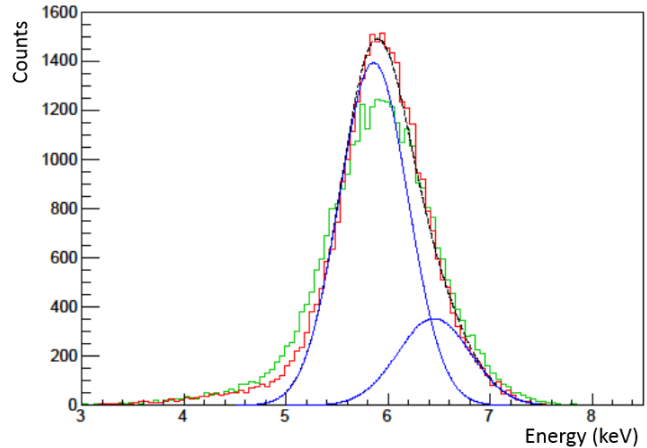

Figure 5.The $^{55}$Fe spectrum obtained with one of the channel. The green histogram is the raw data and the red one is a data removed common mode noise from raw data. The blue lines are the best fit at Mn Kα 5.9 keV and Mn kβ 6.5 keV.

## 4. RADIATION HARDNESS OF ALEX SERIES

**4.1 Specifications of the Radiation Tests**

We test the both Total Ionizing Dose (TID) effect, and Single Event Effect (SEE) especially focusing on the single event upset (SEU). TID is the accumulating damage mainly due to the geo-magnetically trapped protons and electrons. The SEU is caused by mainly two kinds of different sources, i.e. protons or heavy ions. In the case of heavy ions, SEU occurs when the energy deposition exceeds a threshold defined by the linear energy transfer (LET). In contrast, most protons pass through the device with little effect, but very few fraction of protons directly react with silicon nuclei and the products of the p+Si reactions occurs SEU. The ALEX has dual inter locked cell (DICE) type of flip-flops which are tolerant of SEU and the LET threshold of SEU reaches a few MeV-cm$^2$/mg[13].

We performed several times of particle radiation tests with a synchrotron beamline at the Wakasa-Wan Energy Research Center (WERC) to investigate the radiation tolerance of the ALEX series. We used carbon, helium and proton beam as shown in Table 4, and the LET calculated with the SRIM-2013[14] is also listed. The beam profile, is measured by the imaging plate placed at the same position in the atmosphere, has a two-dimensional normal distribution. The beam flux is measured by the Faraday cup inside the beamline. The beam is irradiated at intervals of 2 sec and the pulse duration is about 0.5 sec. The DICE flip-flops are monitoring their high/low state each other, and create one-bit SEU signal if the comparator detects the difference of their state. In these experiments, we measured the SEU rate (cross section of SEU) and the increase of consumption current. Although the ALEX-02 is slightly different from the ALEX-01 in circuit configurations and the number of flip-flop (2 % increasing), we considered that two ASICs are equivalent radiation tolerance and comprehensively investigated.

---
[2] The strip width of SSD installed in Kanazawa-SAT$^3$ is 150 μm.

Table 4. Specifications of the beams at WERC

| Species | Kinetic Energy [MeV] | LET [MeV-cm$^2$/mg]] | Beam Width [cm] | Target |
|---|---|---|---|---|
| Carbon | 200 | 2.3 | 0.83 | ALEX-01 #03 |
| Helium | 220 | $3.6 \times 10^{-2}$ | 0.91 | ALEX-01 #10 |
|  | 220 | $3.6 \times 10^{-2}$ | 0.91 | ALEX-01 #01 |
| proton | 200 | $3.6 \times 10^{-3}$ | 0.93 | ALEX-02 #03 |
|  | 200 | $3.6 \times 10^{-3}$ | 0.90 | ALEX-02 #02 |

### 4.2 Results of the Radiation Tests

We observed two SEU signals for the irradiation of $6.0 \times 10^{11}$ carbon ions, one time for $6.2 \times 10^{10}$ helium ions, six times for $7.8 \times 10^{10}$ protons, and the SEU did not occur for $7.7 \times 10^{10}$ helium ions, $7.8 \times 10^{10}$ protons. Figure 6 (left) shows the cross section of the SEU ($\sigma_{SEU}$) calculated by these results. The filled squares show the data with SEU detections, and the arrows show the upper limits of $\sigma_{SEU}$ in 90% confidence level estimated by assuming the Poisson statistics for all data without SEU. In general, SEU probability are described by the Weibull function, and the $\sigma_{SEU}$ increases rapidly when the LET of the incident particle exceeds the threshold level[15]. Since the $\sigma_{SEU}$ is almost constant with LET in Figure 6 (left), it is considered that the LET threshold of SEU is higher than a few MeV-cm$^2$/mg and SEU might be occurred by the products of the nuclear reactions. According to those experiments, we conclude SEU occurs once for $10^{11}$ particle irradiations independent of particle nuclear species.

The orbit of the HiZ-GUNDAM is sun synchronous orbit with a local time of 9 and 21 hours. The geo-magnetically trapped protons in the South Atlantic Anomaly are predominant sources in the orbit. Then, it is assumed that the flux of protons is about $3 \times 10^9$ particles/cm$^2$/year and that the equivalent dose rate is several kilorad/year[10]. Thus we concluded that ALEX-02 is enough tolerant for the cosmic-ray particles. From now on, we must perform SEU test with a particle exceed 2.3 MeV-cm$^2$/mg with heavier ions (e.g. iron ions), and estimate the SEU rate with a particle exceed the threshold.

Figure 6 (right) shows the measured consumption current as a function of TID. One can see the current in the digital circuits rapidly increase over the TID of 20 krad irradiation for silicon, and reach 0.3 mA at 70 krad. We observed that the analog voltage outputs set by DACs also changed synchronizing with the increasing of digital currents. Although these experiments are in high doze rate condition, we may say that the digital circuit of ALEX is relatively weaker than the SEU effects.

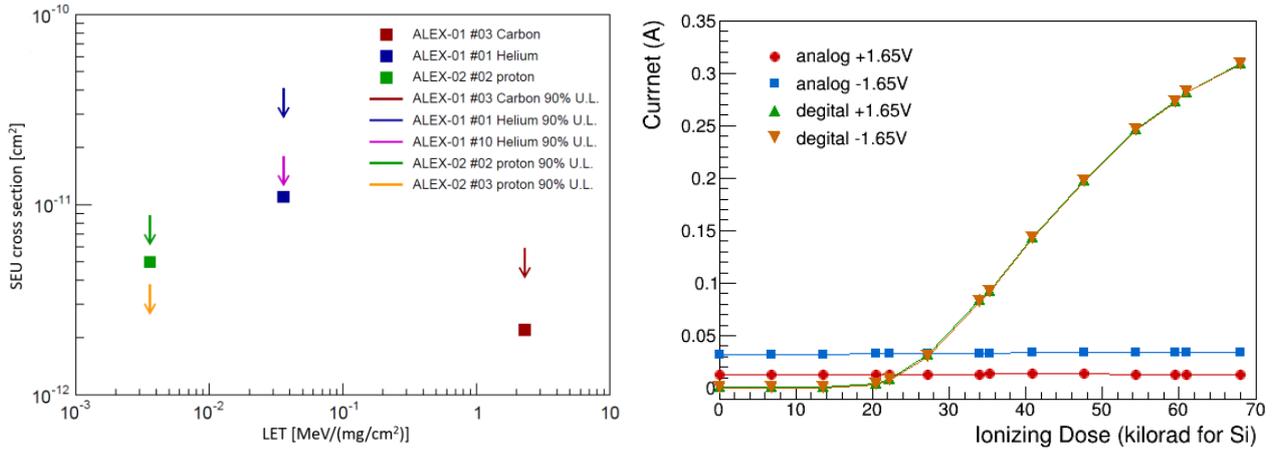

Figure 6. (Left) The cross section of $\sigma_{SEU}$ calculated by result of the test. The filled squares and arrows show the results of SEU detections and the 90 % upper limits without SEU calculated with Poisson statistics. (Right) The measured consumption current as a function of TID.

## 5. ENGINEERING MODEL

We developed an engineering model of two sets of 1-dimensional X-ray imaging detector. Figure 7 (left) shows the sensor board, which contains a large SSD with 512 channel strips (256 ch × 2 lines) and 8 readout ASICs (ALEX-02). The strip pitch is 300 μm with 150 μm width of electrodes. The effective area is 76.8 mm × 32.0 mm for each sensor (24.6 cm$^2$ × 2, i.e. 1/20 scale of final configuration). Figure 7 (right) shows a digital electronics board which has two field programmable gate arrays (FPGAs). One of the FPGA controls and readouts 8 ASICs in parallel, and also makes histograms of X-ray lightcurve, spectrum and images. We burned an 8051 IP core into another FPGA and created mission CPU as telemetry and command interface between satellite's main CPU. In Figure 8, we show a complete settings of EM with X- and Y-dimensional sensor boards. We have one mission CPU which controls two FPGAs on each digital board. All integration circuits installed on the digital board are equivalent to the radiation tolerant devices (pin compatible consumer products). However, several peripheral circuit devices are fully consumer products we selected by the beam irradiation experiments at WERC. We are now performing several environmental test (i.e. the thermal cycle test, vacuum test and the vibration test) by using EM.

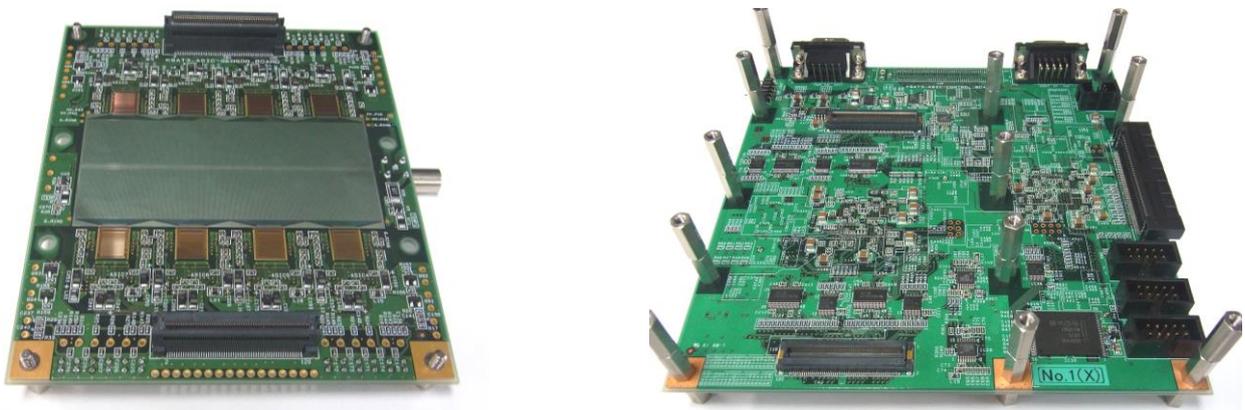

Figure 7. The sensor board (left) and the control board (right) of the engineering model.

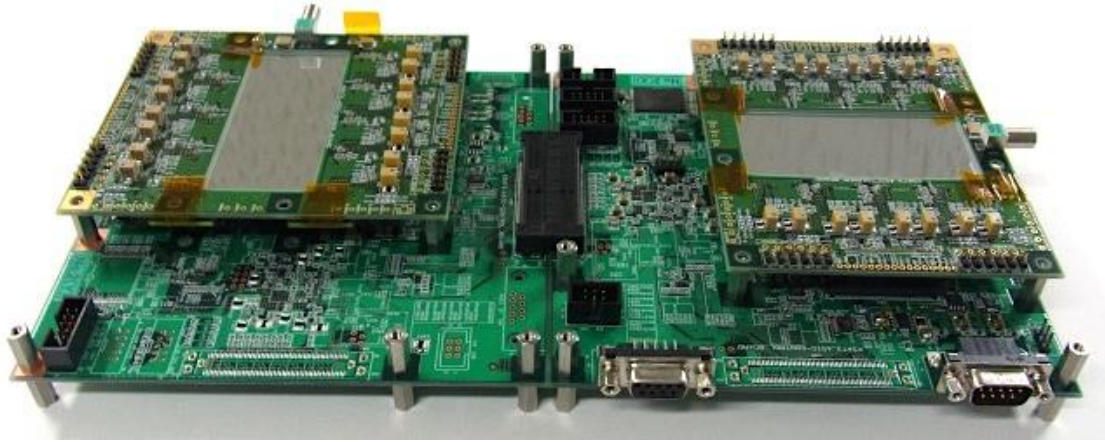

Figure 8. The complete setting of EM with X- and Y-dimensional sensor board.

## 6. SUMMARY

We are developing the wide-field low-energy X-ray imaging detector for HiZ-GUNDAM mission. Especially the high gain readout ASIC is key device for covering the energy range of 1-20 keV. The energy threshold of the first version of ASIC is not satisfied the mission requirement, so we developed the second version of ASIC, named "ALEX-02". The ALEX-02 has longer time constant than one of the first version and its energy threshold is 1.5 keV with no-load condition. We confirmed that we accepted the variation in the characteristics of 64 channels (i.e. baseline voltage, gain or linearity and energy resolution) in the ALEX-02. The beam radiation test was performed with carbon, helium and proton to investigate the radiation hardness of ALEX series. The ALEX can work at least a few years and SEU occurs only one time in about one hundred years. We developed the engineering model of the 1-direction X-ray imaging unit. Using the engineering model, we develop the middleware and perform environmental test.